      [1999/12/01 v1.4c Il Nuovo Cimento]
\documentclass{cimento}

\usepackage{graphicx}
\usepackage{amsmath}
\usepackage{amssymb}
\usepackage{mathrsfs}
\usepackage{dsfont}
\usepackage{epsfig}
\usepackage{graphicx}
\usepackage{subfigure}

\newcommand{\beq}{\begin{equation}}
\newcommand{\eeq}{\end{equation}}
\newcommand{\ba}{\begin{array}}
\newcommand{\ea}{\end{array}}
\newcommand{\bea}{\begin{eqnarray}}
\newcommand{\eea}{\end{eqnarray}}

\def\cO{{\cal O}}

\title{Theoretical Models for Neutrino Masses}
\author{L.~Merlo\from{dep}\from{ias}\thanks{Email: luca.merlo@ph.tum.de}}
\instlist{\inst{dep}Physik-Department, Technische Universit\"at M\"unchen, 
James-Franck-Strasse, D-85748 Garching, Germany
\inst{ias}
TUM Institute for Advanced Study, Technische Universit\"at M\"unchen,
Lichtenbergstrasse 2a, D-85748 Garching, Germany}
\PACSes{
\PACSit{14.60.Pq}{Neutrino mass and mixing}
\PACSit{11.30.Hv}{Flavor symmetries}
\PACSit{12.60.Jv}{Supersymmetry}}
\begin{document}

\maketitle

\begin{abstract}
The recent measurements of the neutrino reactor angle require a re-examination of flavour models based on discrete groups. Indeed, when these models deal with the Tri-Bimaximal, the Bimaximal and the Golden Ratio mixing patterns, some tensions arise in order to accommodate the reactor angle. In particular, strong constraints come from lepton flavour violating processes, like $\mu \to e \gamma$. We present the analysis and the main results.
\end{abstract}

\section{Neutrino Data and Predictive Mass Patterns}

Solar and atmospheric neutrino experiments have established the appearance and the disappearance of specific flavour neutrinos, that finds the best explanation in the oscillation of active neutrinos. Even if the issue of the presence of one or more sterile neutrinos must still be clarified, global fits with only three oscillating active neutrinos well reproduce the data. Nowadays, the most recent results\footnote{Notice that the best fit values for the reactor angle differ for a factor of 2 in the two fits, due to the exclusion (Fogli {\it et al.}) or the inclusion (Schwetz {\it et al.}) of the data from SBL neutrino experiments with a baseline $<100$ m.} on the oscillation data can be summarized in Tab.~\ref{tab:data}.

More recently, T2K data \cite{Abe:2011sj} showed evidences for a non-vanishing reactor angle at the $3\sigma$ level. Subsequently also MINOS \cite{Adamson:2011qu} and Double Chooz \cite{Abe:2011fz} presented their results, in agreement with T2K one. In the last months the Daya Bay \cite{An:2012eh} and RENO \cite{Ahn:2012nd} experiments have released their results on the observation of electron anti-neutrino disappearance, providing at more than $5\sigma$ and $6\sigma$, respectively, the evidence for a non-vanishing reactor angle:
\beq
\sin^2\theta_{13}=0.024 \pm 0.005\qquad \text{[Daya Bay]}\,,\qquad
\sin^2\theta_{13}=0.029 \pm 0.006\qquad \text{[RENO]}\,.
\eeq
The average of the results for the reactor angle from the cited experiments, for normal (inverted) hierarchy, is given by
\beq
\sin^2\theta_{13}=0.022\pm0.004\,(0.023\pm0.004)\,.
\label{OurReactor}
\eeq

\begin{table}[h]
\vspace{-0.3cm}
\caption{Fits to neutrino oscillation data. For the fit of Schwetz {\it et all} \cite{Schwetz:2011zk}, different results have been found for the two hierarchies: the IH is shown in the brackets. In both the fits, the results correspond to the new reactor fluxes, accounting for the T2K \cite{Abe:2011sj} and MINOS \cite{Adamson:2011qu} data on the reactor angle.} 
\label{tab:data}
\begin{tabular}{ccc}
  \hline
  &&\\[-3mm]
  			 														& Fogli {\it et al.} \cite{Fogli:2011qn} 	& Schwetz {\it et al.} \cite{Schwetz:2011zk} \\[1mm]
  \hline
  &&\\[-2mm]
  $\Delta m^2_{sun}~(10^{-5}~{\rm eV}^2)$ 	& $7.58^{+0.22}_{-0.26}$ 					& $7.59^{+0.20}_{-0.18}$  \\[1mm]
  $\Delta m^2_{atm}~(10^{-3}~{\rm eV}^2)$ 	& $2.35^{+0.12}_{-0.09}$ 					& $2.50^{+0.09}_{-0.16}\,[2.40^{+0.08}_{-0.09}]$  \\[1mm]	
  $\sin^2\theta_{12}$ 									& $0.312^{+0.017}_{-0.016}$				& $0.312^{+0.017}_{-0.015}$ \\[1mm]
  $\sin^2\theta_{23}$ 									& $0.42^{+0.08}_{-0.03}$ 					& $0.52^{+0.06}_{-0.07}\,[0.52\pm0.06]$ \\[1mm]
  $\sin^2\theta_{13}$ 									& $0.025\pm0.007$ 							& $0.013^{+0.007}_{-0.005}[0.016^{+0.008}_{-0.006}]$  \\[1mm]
  \hline
\end{tabular}
\end{table}

From the theoretical side, a great effort has been put to construct flavour models that are able to describe and explain the experimental results. Before the new data on the reactor angle, the attention was focussed on a particular class of mixing patterns, for their high predictive power. In the following we will concentrate of the Tri-Bimaximal \cite{Harrison:2002er,Xing:2002sw} (TB), the Golden Ratio \cite{Kajiyama:2007gx,Rodejohann:2008ir,Adulpravitchai:2009bg,Everett:2008et,Feruglio:2011qq} (GR) and the Bimaximal \cite{Vissani:1997pa,Barger:1998ta,Nomura:1998gm,Altarelli:1998sr} (BM) schemes. All these mixing schemes predict a maximal atmospheric angle and a vanishing reactor angle,
\beq
\sin^2\theta_{23}=\dfrac{1}{2}\,,\qquad\qquad
\sin^2\theta_{13}=0\,,
\label{OtherPredictions}
\eeq
while they differ for the prediction of the solar angle:
\beq
\sin^2\theta^{TB}_{12}=\dfrac{1}{3}\,,\qquad\qquad
\sin^2\theta^{GR}_{12}=\dfrac{2}{5+\sqrt5}\equiv\dfrac{1}{\sqrt5 \phi}\,,\qquad\qquad
\sin^2\theta^{BM}_{12}=\dfrac{1}{2}\,.
\label{SolarPredictions}
\eeq
Considering the predicted value of the solar angle for the three mixing schemes, while the TB and the GR patterns agree well with the data, the BM one does not at more than $5\sigma$. The interest on the BM pattern is mainly due to its relation with the so-called Quark-Lepton complementarity \cite{Altarelli:2004jb,Raidal:2004iw,Minakata:2004xt} (QLC): the QLC consists in a numerical relation such that the sum of the experimental values of the lepton solar angle and of the Cabibbo angle is roughly $\pi/4$. From here the idea to revert this expression and write $\theta^{exp}_{12}\approx\theta^{BM}_{12}-\theta_C$, where the BM prediction for the solar angle enters \cite{Altarelli:2009gn,Toorop:2010yh,Meloni:2011fx}.

Entering more in details of the these predictive patterns, the unitary matrices corresponding to the mixing angles listed in eqs.~(\ref{OtherPredictions}) and (\ref{SolarPredictions}) are the following:
\beq
\begin{gathered}
U_{TB}=\left (\begin{array}{ccc} 2/\sqrt{6} & 1/\sqrt{3} & 0\\
-1/\sqrt{6} & 1/\sqrt{3} & -1/\sqrt{2}\\
-1/\sqrt{6} & 1/\sqrt{3} & 1/\sqrt{2}\\
\end{array}\right) \,,\qquad
U_{BM}=\left (\begin{array}{ccc} 1/\sqrt{2} & -1/\sqrt{2} & 0\\
1/2 & 1/2 & -1/\sqrt{2}\\
1/2 & 1/2 & 1/\sqrt{2}\\
\end{array}\right)\,,\\
U_{GR}=\left (\begin{array}{ccc} \cos\theta^{GR}_{12} & \sin\theta^{GR}_{12} & 0\\
\sin\theta^{GR}_{12}/\sqrt{2} & -\cos\theta^{GR}_{12}/\sqrt{2} & 1/\sqrt{2}\\
\sin\theta^{GR}_{12}/\sqrt{2} & -\cos\theta^{GR}_{12}/\sqrt{2} & -1/\sqrt{2}\\
\end{array}\right)\,.
\end{gathered}
\eeq
In all the three mixing matrices the $13$ entry is zero, corresponding to a vanishing reactor angle and to an undetermined Dirac CP phase. Moreover, all the entries are pure numbers and ensure the independence of the mixing angles from the specific neutrino spectrum: this feature is commonly linked to neutrino mass matrices that are form-diagonalizable \cite{Low:2003dz} (FD).

\section{Neutrino Flavour Models and the Reactor Angle}

The predictive patterns described in the previous section have been considered as a starting point to reproduce the experimental data. To this aim discrete non-Abelian flavour symmetries are extremely successful and have been implemented in different approaches. In the following we will concentrate on models where the flavour symmetry is: global, in order to avoid the presence of a new force, the corresponding gauge bosons and their flavour violating effects \cite{Grinstein:2010ve,Buras:2011zb,Buras:2011wi} (a gauged discrete symmetry should be considered as a remnant of a gauged continuous symmetry breaking); spontaneously broken at the high-energy, in order to prevent strong flavour violating effects common in the low-energy flavour breaking mechanism \cite{Toorop:2010ex,Toorop:2010kt,Branco:2011iw}; broken by a set of scalar fields, called flavons, that transform only under the flavour symmetry, for which a well-defined vacuum alignment mechanism can be constructed (this is one of the main advantages with respect to continuous symmetries \cite{Feldmann:2009dc,Alonso:2011yg}). In models that fulfill the previous description, the Yukawa Lagrangian is usually written in terms of non-renormalizable operators \cite{Varzielas:2010mp} suppressed by suitable powers of the cut-off scale $\Lambda_f\approx\Lambda_L\approx\Lambda_{GUT}$, where $\Lambda_L$ is the scale of lepton number violation and $\Lambda_{GUT}$ the GUT scale:
\beq
\mathcal{L}_Y=\dfrac{\left(Y_{e}[\varphi^n]\right)_{ij}}{\Lambda_f^n}\,e^c_i\,H^\dag\,\ell_j+\dfrac{\left(Y_\nu[\varphi^m]\right)_{ij}}{\Lambda_f^m}\dfrac{(\ell_i\,\tilde H^*)(\tilde H^\dag\,\ell_j)}{2\Lambda_L}\,.
\eeq
Here the Weinberg operators describes the neutrinos, but a completely similar Yukawa Lagrangian can be written for the See-Saw mechanisms. When the flavour and the electroweak symmetries are broken, the charged lepton and neutrino mass matrices are generated. In these models, considering only the lowest dimensional operators, the TB, GR and BM patterns could naturally arise as the lepton mixing matrix $U_{PMNS}$. Considering also the higher dimensional operators, new contributions correct the LO PMNS matrix and are responsible for deviations from the TB, GR and BM predicted mixing angles. 

The main ingredient that allows to recover these mixing patterns and their corrections is the flavour breaking mechanism: the flavons develop vacuum expectation values (VEVs) in specific directions of the flavour space, such that the starting flavour symmetry $G_f$ is broken down to two distinct subgroup, $G_\nu$ and $G_\ell$ in the neutrino and charged lepton sectors, respectively. We will indicate the set of flavons that lead to $G_\nu$ ($G_{\ell}$) as $\Phi_\nu$ ($\Phi_\ell$). $G_\nu$ and $G_\ell$ represent the low energy symmetries of the neutrino and charged lepton mass matrices: some examples are $G_\nu=Z_2\times Z_2$\footnote{In some cases, $G_f$ it broken down to $G_\nu=Z_2$, but an additional accidental $Z_2$ symmetry is also present in this sector \cite{Toorop:2011jn,deAdelhartToorop:2011re,Ge:2011ih,Hernandez:2012ra}.} and $G_\ell=Z_n$, with $n>3$.

Furthermore, the existence of such symmetry breaking mechanism is usually enforced in a supersymmetric context, even if other possibilities have been studied \cite{Altarelli:2005yp,Bazzocchi:2008ej}: in the following we will consider only supersymmetry flavour models.

Whether the final PMNS reproduces the experimental data depends on specific features of the models: in the following we identify three major classes that well represent the present situation in model building \cite{Altarelli:2012bn,Altarelli:2012ss,Bazzocchi:2012st}. The GR models can be associate with the TB ones for what concerns the results of the present analysis.

\subsection{Typical $A_4$ Models for the TB Mixing Pattern}

For this class, we consider for definiteness the model in Refs.~\cite{Altarelli:2005yp,Altarelli:2005yx,Altarelli:2009kr}, but the analysis applies to a broader range of models based on $A_4$ (see Ref.~\cite{Altarelli:2012bn,Altarelli:2012ss,Bazzocchi:2012st} for details) or on other symmetries (\textit{i.e.} Refs.~\cite{Feruglio:2007uu,Bazzocchi:2009pv,Bazzocchi:2009da}). The neutrino and charged lepton mass matrices can be written as
\beq
m_e=m_e^{(0)}+\delta m_e^{(1)}\,,\qquad\qquad
m_\nu=m_\nu^{(0)}+\delta m_\nu^{(1)}\,,
\eeq
where $m_e^{(0)}=\text{diag}(y_e,\,y_\mu,\,y_\tau)\,v_d\,\eta$, with $v_d$ the VEV of $H_d$ and $\eta=\langle\Phi_\ell\rangle/\Lambda_f$ a small parameter that breaks $A_4$ down to $G_\ell$, and $m_\nu^{(0)}$ is diagonalized by the TB mixing matrix. 

\begin{figure}[h!]
 \centering
   \subfigure[{Correlation with $\sin^2\theta_{12}$}]
   {\includegraphics[width=6.5cm]{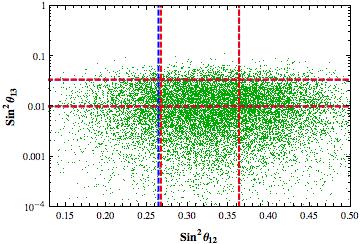}}
   \subfigure[{Correlation with $\sin^2\theta_{23}$}]
   {\includegraphics[width=6.5cm]{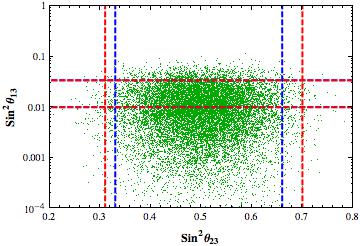}}
 \caption{ {\bf Typical {\boldmath$A_4$\unboldmath} Models.} $\sin^2\theta_{13}$ as a function of $\sin^2\theta_{12}$ ($\sin^2\theta_{23}$) is plotted on the left (right), following eq.~(\ref{sinNLOTB}). The vertical lines represent the $3\sigma$ values for $\sin^2\theta_{12}$ and $\sin^2\theta_{23}$, following the Fogli {\it et al.} \cite{Fogli:2011qn} fit, in blue, and the Schwetz {\it et al.} \cite{Schwetz:2011zk} fit, in red. The horizontal lines refers to the $3\sigma$ values for $\sin^2\theta_{13}$ as in eq.~(\ref{OurReactor}).}
\vspace{-0.3cm}
\label{fig:Sin12qvsSin13qTB_NIH}
\end{figure}

In a typical model, the NLO contributions to both the mass matrices correct all the entries and are of the same order of magnitude, that we can parametrise with $\xi=\langle\Phi_\ell\rangle/\Lambda_f\approx\langle\Phi_\nu\rangle/\Lambda_f$, a small parameter that 
breaks also the subgroups $G_\ell$ and $G_\nu$. In this case, the mixing angles receive deviations from the initial TB values and we can write:
\beq
\begin{aligned}
\sin^2\theta_{23}&=\frac{1}{2}+{\cal R}e(c^e_{23})\,\xi+\dfrac{1}{\sqrt{3}}\left({\cal R}e(c^\nu_{13})-\sqrt2\,{\cal R}e(c^\nu_{23})\right)\,\xi\\
\sin^2\theta_{12}&=\frac{1}{3}-\frac{2}{3}{\cal R}e(c^e_{12}+c^e_{13})\,\xi+\dfrac{2\sqrt2}{3}\,{\cal R}e(c^\nu_{12})\,\xi\\[1mm]
\sin\theta_{13}&=\dfrac{1}{6}\left|3\sqrt2\left(c^e_{12}-c^e_{13}\right)+2\sqrt3\left(\sqrt2\,c^\nu_{13}+c^\nu_{23}\right)\right|\,\xi\,.
\end{aligned}
\label{sinNLOTB}
\eeq
where $c^{e,\nu}_{ij}$, complex random number with absolute value of order 1, is the $ij$ entry of unitary matrices that diagonalize the charged lepton and neutrino mass matrices at the NLO. Accordingly with these expressions, the success rate to reproduce all the three mixing angles inside the corresponding $3\sigma$ ranges is maximized for $\xi=0.07$ for both the NH and IH.
We analyze quantitatively the expressions in eq.~(\ref{sinNLOTB}) and their correlations in Fig.~\ref{fig:Sin12qvsSin13qTB_NIH}. The $c^{e,\nu}_{ij}$ parameters are treated as complex complex numbers with absolute values following a Gaussian distribution around 1 with variance 0.5. In the plots we show only the NH case. The IH case is similar.

As we can see, the plots are representing the general behaviour of this class of models: $\sin^2\theta_{13}$ increases with $\xi$, but correspondingly also the deviation of $\sin^2\theta_{12}$ from $1/3$ does. As a result, even for the value of $\xi$ that maximizes the success rate, the requirement for having a reactor angle inside its $3\sigma$ error range corresponds to a prediction for the solar angle that is no more in good agreement with data.

\subsection{Special $A_4$ Models for the TB Mixing Pattern}

There are some special models based on the group $A_4$ \cite{Lin:2009bw}, in which the LO predictions for the mass matrices are the same as in the previous section, but the corrections are not completely generic. In the specific case of the model in Ref.~\cite{Lin:2009bw}, the charged lepton mass matrix still receive generic corrections proportional to $\xi=\langle\Phi_\ell\rangle/\Lambda_f$, but the neutrino mass matrix is corrected only in determined directions: the unitary matrix that digitalize the final neutrino mass matrix is given by
\beq
U_\nu=U_{TB}\,V\,,\qquad\qquad
\text{with}\qquad\qquad
V=\left(
        \begin{array}{ccc}
           \alpha  & 0 & \xi' \\
            0 & 1 & 0 \\
            -\xi'^* & 0 & \alpha^* \\
        \end{array}
\right)\,,
\eeq
where $|\alpha|^2+|\xi'|^2=1$, where $\xi'=\langle\Phi_\nu\rangle/\Lambda_f$. In this specific model $\xi'>\xi$. The final expressions for the neutrino mixing angles after the inclusion of all these corrections are given by:
\bea
\sin\theta_{13}&=&\left|\sqrt{\frac{2}{3}}\,\xi'+\frac{c^e_{12}-c^e_{13}}{\sqrt{2}}\,\xi\right|\,,\qquad\qquad\qquad
\delta\approx\arg\xi'\,,\\
\sin^2\theta_{12}&=&\frac{1}{3}+\frac{2}{9}\,|\xi'|^2-\frac{2}{3}\,{\cal R}e(c^e_{12}+c^e_{13})\,\xi\,,
\label{sinNNLOTBLinSol}\\
\sin^2\theta_{23}&=&\frac{1}{2}+\frac{1}{\sqrt{3}}\,|\xi'|\,\cos\delta+{\cal R}e(c^e_{23})\,\xi\,.
\label{sinNNLOTBLinAtm}
\eea

\begin{figure}[h!]
 \centering
   \subfigure[Correlation with $\sin^2\theta_{12}$]
   {\includegraphics[width=6.5cm]{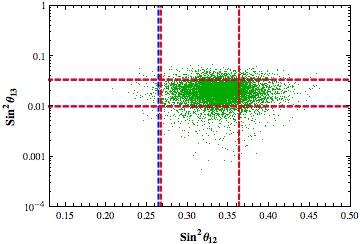}}
   \subfigure[Correlation with $\sin^2\theta_{23}$]
   {\includegraphics[width=6.5cm]{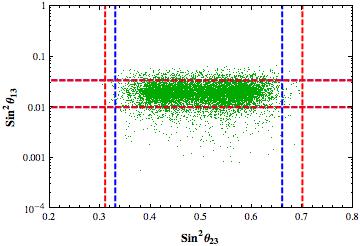}}
 \caption{ {\bf Special {\boldmath$A_4$\unboldmath} Models.} $\sin^2\theta_{13}$ as a function of $\sin^2\theta_{12}$ ($\sin^2\theta_{23}$) is plotted on the left (right), following eqs.~(\ref{sinNNLOTBLinSol}) and (\ref{sinNNLOTBLinAtm}). The vertical and the horizontal lines are as in fig.~\ref{fig:Sin12qvsSin13qTB_NIH}.}
 \label{fig:Sin23qvsSin13qTBLin_NIH}
\vspace{-0.3cm}
\end{figure}

The success rate to reproduce all the three mixing angles inside their corresponding $3\sigma$ error ranges is maximized by $|\xi'|=0.166 (0.171)$ for the NH (IH). The parameters have been chosen such that $\xi$ is a real number in $[0.005,\,0.06]$ and $c^e_{ij}$ are random complex numbers with absolute values following a Gaussian distribution around 1 with variance 0.5. We analyze quantitatively the deviations in eqs.~(\ref{sinNNLOTBLinSol}) and (\ref{sinNNLOTBLinAtm}) and their correlations in Fig.~\ref{fig:Sin23qvsSin13qTBLin_NIH}: in the plots on the left (right) column, we show the correlations in eqs.~(\ref{sinNNLOTBLinSol}) and (\ref{sinNNLOTBLinAtm}) between $\sin^2\theta_{13}$ and $\sin^2\theta_{12}$ or $\sin^2\theta_{23}$, respectively: $\xi'$ is a complex number with absolute values equal to $0.166$. In the plots we show only the NH case. The IH case is similar. For this choice of the parameters, the model can well describe all three angles inside the corresponding $3\sigma$ interval, and its success rate is much larger than that of the typical TB models.

\subsection{$S_4$ Models for the BM Mixing Pattern}

\begin{figure}[h!]
 \centering
  \subfigure[{Correlation with $\sin^2\theta_{12}$.}]
   {\includegraphics[width=6.5cm]{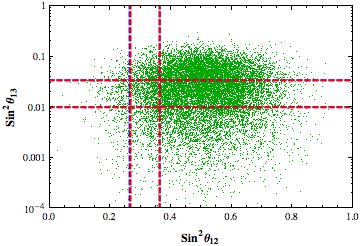}}
  \subfigure[{Correlation with $\sin^2\theta_{12}$ with $c_{13}=0$.}]
   {\includegraphics[width=6.5cm]{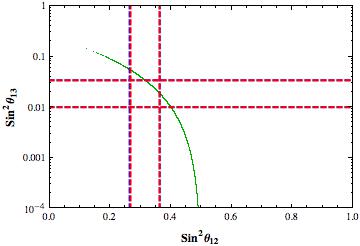}}
 \caption{{\bf {\boldmath$S_4$\unboldmath} Models.} $\sin^2\theta_{13}$ as a function of $\sin^2\theta_{12}$ is plotted, following eq.~(\ref{sinNLOBM}).}
 \vspace{-0.5cm}
 \label{fig:Sin12qvsSin13qBM_NIH}
\end{figure}

For the last class, we focus on a representative model based on the $S_4$ discrete group \cite{Altarelli:2009gn}. In this case, $m_e^{(0)}$ is still the diagonal matrix with the charged lepton masses, but $m_\nu^{(0)}$ is diagonalized by the BM mixing matrix. At the higher orders, the neutrino mass matrix preserves the same LO flavour structure up to the NNLO level. On the contrary, the changed lepton mass matrix is corrected at the NLO in all the entries, but not in the $23$ and $32$ ones. As a result, the final neutrino mixing angles at the NLO are given by:
\beq
\sin\theta_{13}		=\frac{1}{\sqrt{2}}\,{|c^e_{12}-c^e_{13}|}\,\xi\,,\qquad
\sin^2\theta_{12}	=\frac{1}{2}-\frac{1}{\sqrt{2}}\,{\cal R}e(c^e_{12}+c^e_{13})\,\xi\,,\qquad
\sin^2\theta_{23}	=\frac{1}{2}\,.
\label{sinNLOBM}
\eeq
To properly correct the BM value of the solar angle to agree with the data, $\xi$ is expected to be $\cO(\lambda_C)$. Studying the success rate to have all the three mixing angles inside the corresponding $3\sigma$ ranges, we find that it is maximized for both the NH and IH when $\xi=0.163$. We analyze quantitatively the expressions in eq.~(\ref{sinNLOBM}) and their predictions in Fig.~\ref{fig:Sin12qvsSin13qBM_NIH}, where $c_{12,13}$ have been taken as random complex numbers with absolute value following a Gaussian distribution around 1 with variance 0.5, while $\xi=0.185(0.194)$. A value close $\cos \delta^{CP}$ close to $-1$ is favoured in order to maximize the success rate \cite{Altarelli:2012bn,Altarelli:2012ss,Bazzocchi:2012st}. In Fig.~\ref{fig:Sin12qvsSin13qBM_NIH}, only for the NH case is shown. The IH case is similar.

Considering the results for the success rates of all the three classes of models, these $S_4$ models are strongly disfavoured with respect to the special $S_4$ ones, while are comparable with respect to the typical $A_4$ models.

\section{Conclusions}

Discrete symmetries can well accommodate the neutrino mixing pattern, especially considering an approach in which the PMNS matrix is given in first approximation by the TB, the GR or the BM patterns. With the new results on the reactor angle, however, it appears suspicious that one of these mixing schemes could be a fundamental structure of nature, while it is getting stronger the feeling that they are simply numerical accidents. Indeed, the type and the size of the corrections necessary to bring these mixing patterns in agreement with the data put sever doubts on their naturalness.

Furthermore, as discussed in a series of papers \cite{Feruglio:2008ht,Ishimori:2008au,Ishimori:2009ew,Feruglio:2009iu,Feruglio:2009hu,Merlo:2011hw} and updated in Ref.~\cite{Altarelli:2012bn,Altarelli:2012ss,Bazzocchi:2012st}, the analysis of lepton flavour violating transitions is fundamental to test flavour models. In particular, with the new results on the reactor angle and the large size of the NLO corrections, the bounds on the supersymmetric parameters space coming from the $\ell_i\to\ell_j\gamma$ decays are strong, even for small $\tan\beta$, and if light supersymmetric particles are found then these models are disfavoured.  Moreover, it appears impossible to satisfy the MEG bound and, at the same time, to reproduce the muon $g-2$ discrepancy.

Even though the huge effort of these years in constructing flavour models to describe masses and mixings for the neutrinos, and more in general for all the fermions, it is discouraging that no illuminating strategy arise form this scenario. On the other hand, this is partially related to the large uncertainties still present in the flavour sector. The hope is that with a better determination of the lepton mixing angles and with the knowledge of the CP phases, the neutrino mass scale, the type of the neutrino nature and spectrum, it will be finally possible to shed light on the origin of the fermion masses and mixings.

\acknowledgments
I warmly thank Guido Altarelli, Ferruccio Feruglio and Emmanuel Stamou for the fruitful collaboration that leaded to the results presented in this talk. I recognise that this work has been partly supported by the TUM -- IAS, funded by the German Excellence Initiative. Finally I thank the organizers of the La Thuile conference for the kind invitation and for their efforts in organizing this enjoyable meeting.


\end{document}